\documentclass{article}
\usepackage{spconf,amsmath,epsfig}
\usepackage{tikz}
\usepackage{adjustbox}
\usepackage{amssymb}
\usepackage{color,soul}
\usepackage[colorlinks, allcolors=black]{hyperref}
\newcommand{\sectionref}[1]{Section \ref{#1}}

\title{A multimodal supervised machine learning approach for satellite-based wildfire identification in Europe}
\name{Angelica Urbanelli$^{1*}$, Luca Barco$^{1*}$, Edoardo Arnaudo$^{1,2}$, Claudio Rossi$^1$
    \thanks{
    This work was carried out in the context of the H2020 projects: SAFERS
(GA n.869353) and OVERWATCH (GA n.101082320) \\
    * Equal contribution.
    }
}
\address{
    1. LINKS Foundation, \textit{AI, Data \& Space (ADS)}, Torino (TO), Italy \\
    2. Politecnico di Torino, \textit{Dipartimento di Automatica e Informatica  (DAUIN)}, Torino (TO), Italy
}
\begin{document}
%
\maketitle
\begin{abstract}
The increasing frequency of catastrophic natural events, such as wildfires, calls for the development of rapid and automated wildfire detection systems. In this paper, we propose a wildfire identification solution to improve the accuracy of automated satellite-based hotspot detection systems by leveraging multiple information sources. We cross-reference the thermal anomalies detected by the  Moderate-resolution Imaging Spectroradiometer (MODIS) and the Visible Infrared Imaging Radiometer Suite (VIIRS) hotspot services with the European Forest Fire Information System (EFFIS) database to construct a large-scale hotspot dataset for wildfire-related studies in Europe. Then, we propose a novel multimodal supervised machine learning approach to disambiguate hotspot detections, distinguishing between wildfires and other events. Our methodology includes the use of multimodal data sources, such as the ERSI annual Land Use Land Cover (LULC) and the Copernicus Sentinel-3 data. Experimental results demonstrate the effectiveness of our approach in the task of wildfire identification.
\end{abstract}
\section{Introduction}
\label{sec:intro}
The world is experiencing a surge in catastrophic natural events, including wildfires, floods, and storms, the frequency and magnitude of which are exacerbated by climate change effects.
The rising global average temperature is expected to increase the risk of fires across a wide range of latitudes, leading to more frequent and impacting events and consequently to the release of a significant amount of smoke, carbon dioxide and heat, which further drives up climate change \cite{ipcc2022report}.
Automated wildfire detection systems are crucial to identify fires in their early stage, allowing a prompt response and a consequent reduction of the impacts on human ecosystems, especially in remote areas that are sparsely populated and urbanized. 

In this context, the Moderate-resolution Imaging Spectroradiometer (MODIS), and the Visible Infrared Imaging Radiometer Suite (VIIRS) provide operational services aimed at detecting thermal anomalies. On top of these systems, various solutions have been developed to convert raw measurements into \textit{hotspot detections}, i.e., thermal anomalies that are likely to be caused by fires. Existing \textit{hotspot detections} services implement algorithms that can adapt to different instruments and resolutions, filtering out as many \textit{false positive} as possible. However, the accuracy of such services is not ideal for integrating their outputs in operational chains aimed at the automatic detection of wildfires.

In this work, we propose to tackle this problem through an \textit{hotspot disambiguation} task, introducing a novel supervised machine learning approach leveraging multiple data sources to increase the accuracy of wildfire detection from satellite-based thermal anomalies in Europe. Moreover, we openly release both the code of the proposed approach as well as the large-scale dataset we use for the performance evaluation \footnote{Code and dataset are available at the following URL: \url{https://github.com/links-ads/hotspot-disambiguation}}.

This paper is organized as follows. In \sectionref{sec:relatedworks}, we present a review of the related works in the field of automatic fire detection using satellite observations, while in \sectionref{sec:dataset} we detail the employed data sources and how we construct our large-scale hotspot dataset. In \sectionref{sec:methodology}, we describe the proposed methodology, together with the machine learning models used, while in \sectionref{sec:experiments} we present our experimental settings and results. Lastly, in \sectionref{sec:conclusions}, we draw our conclusions and discuss potential future works.

\section{Related works}
\label{sec:relatedworks}
In the context of fire detection, significant efforts have been carried out to create scalable approaches able to extract thermal anomalies from raw satellite data. These works, carried out by \textit{Giglio et al.} can adapt to different satellites, resolutions, sensors, and data sources, e.g., MODIS and VIIRS \cite{Giglio2003AnEC,GiglioVIIRS}, providing collections of hotspots that can be used for downstream tasks, such as alerting or mapping services. 
Since there is no manual or automatic validation provided with the outputs of these algorithms, some studies compare the output of MODIS and VIIRS hotspot detection algorithms using various data sources, namely Landsat feeds \cite{schroeder2016}, Geostationary Earth Observation satellites with a lower revisit time \cite{engel2020}, the ESA Climate Change Initiative Land Cover and Global Annual Burned Area Maps \cite{fu2020}, and the Copernicus Global Land Service Land Cover \cite{coskuner2022}. While a manual validation approach is not viable nor feasible without relying on data from other satellites or alternative algorithms, the automatic validation of currently detected hotspots for wildfires identification remains an open challenge.

In this work, we leverage on the European Forest Fire Information System (EFFIS), which is the most important European database of burned areas, to extract the MODIS and VIIRS hotspots that correspond to real wildfires and use such ground truth to train different supervised machine learning models using multimodal data, namely the ESRI annual Land Use Land Cover (LULC), Sentinel-3 Sea and Land Surface Temperature Radiometer (SLSTR) and  Ocean and Land Colour Instrument (OLCI). We compare the performance of both classical machine learning models and more recent deep learning architectures, obtaining substantial improvements in wildfire classification accuracy.

\section{Dataset}
\label{sec:dataset}

\subsection{Data sources}
We build our wildfire hotspot dataset by harmonizing several data sources, specifically: MODIS, VIIRS, EFFIS, Sentinel-3 SLSTR and OLCI, LULC.

MODIS is an instrument developed by NASA, collecting data across 36 spectral bands with a spatial resolution varying within the range of 250 to 1000 meters. It is able to identify areas that are hotter than their surroundings and flag them as active fires. In this study, we exploit the MCD14ML data collection, which has a spatial resolution of 1km.
Similarly, VIIRS is a NASA instrument designed to be complementary to MODIS. It uses similar fire detection algorithms but with a higher spatial resolution, specifically VNP14 and VNP14IMG with 750 m/pixel and 350 m/pixel, respectively.

EFFIS instead is the reference European service providing critical information about wildfire risks and impacts, allowing for more effective prevention, mitigation, and response efforts. Since 2015, it is part of the Copernicus Emergency Management System \footnote{www.copernicus.eu}. The EFFIS Burnt Area product contains a list of manually validated wildfires events detected from the daily processing of MODIS and Sentinel-2 satellite imagery, at $250m$ and $20m$ spatial resolution, respectively. Wildfires registered in this database must have a minimum area of 30 hectares, thus excluding smaller fires. In addition, the database also provides a rough estimate of the start and end time for each event, but they may not correspond to the actual timeline, especially considering its end. For this reason, we estimate the extinction date of the fire considering the first date on which there are less than 2 hotspots in that area.

Sentinel-3 is a satellite mission that provides global observations of Earth's oceans, land surfaces, and atmosphere to support environmental and climate monitoring. The satellite was launched on 16/02/2016 and hosts four instruments, including the SLSTR and the OLCI.
The SLSTR instrument has 9 spectral bands and 2 additional active fire bands useful for fire monitoring. It has a spatial resolution of 1 km, with a revisit time of less than half a day.
The OLCI instrument has 21 spectral bands that measure biological information, including the land cover. It has a spatial resolution of 300m with a revisit time of less than 2 days with 2 satellites.
Data are gathered with a spatial resolution of 300m. A bicubic resampling algorithm is used for upsampling the 1 km resolution SLSTR bands.

Lastly, LULC is a collection of annual maps of land cover and land use provided by ESRI, and derived from Sentinel-2 imagery at 10m resolution by exploiting deep learning models. These maps comprise 9 land cover classes, namely: \textit{water}, \textit{trees}, \textit{flooded vegetation}, \textit{crops}, \textit{built area}, \textit{bare ground}, \textit{snow/ice}, \textit{clouds and rangeland} \cite{9553499}.
For this analysis, the map of 2018 has been used with a spatial resolution of 300m to be compliant with Sentinel-3 data.

\begin{figure}[htb]
\begin{minipage}[b]{\linewidth}
  \centering
\centerline{\epsfig{figure=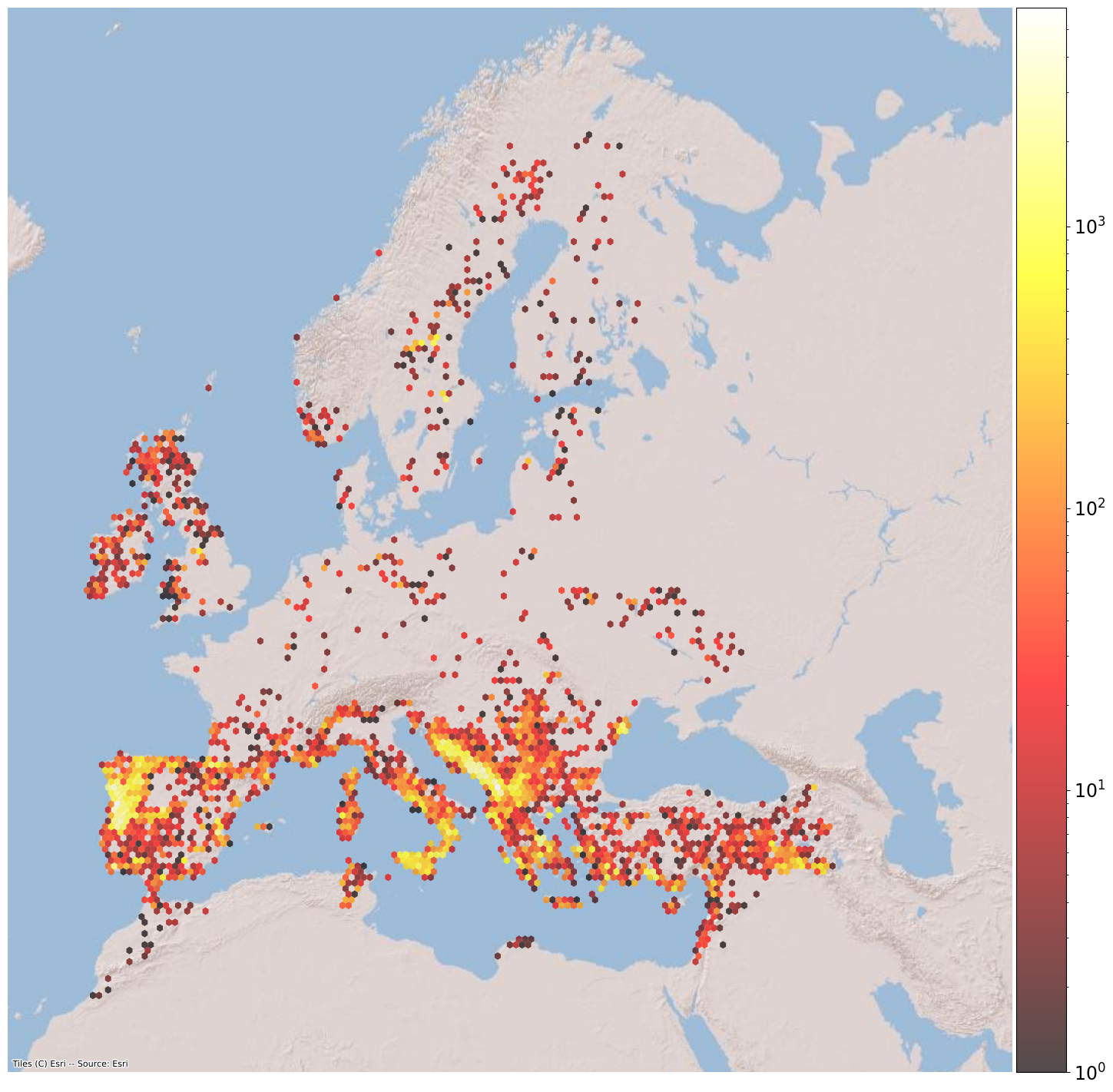,width=.8\linewidth}}
\end{minipage}
\caption{Distribution of hotspots intersecting with EFFIS burned areas. The concentration of points is visually depicted through a logarithmic color scale.}
\label{fig:hotspots}
\end{figure}
\vskip -0.5cm

\begin{figure*}[ht!]
    \centering
    \includegraphics[width=1\textwidth]{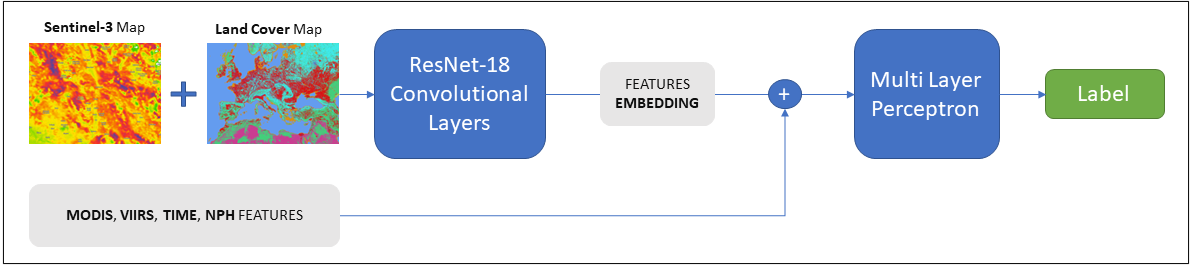}
    \caption{Architecture schema for the end-to-end solution (RN-E2E).}
    \label{fig:E2E}
\end{figure*}

\subsection{Cross-Referencing}
We construct our hotspot dataset by cross-referencing MODIS and VIIRS thermal anomalies extracted from the algorithm by \textit{Giglio et al.} on the Mediterranean region with the EFFIS wildfire database, obtaining a collection of more than 9.5M hotspots from January 2012 to August 2022. Specifically, we consider a hotspot to be positive if it co-occurs in both space and time with the corresponding burned areas, indicating a strong association between the detected thermal anomalies and the documented fire events in the EFFIS database. Figure \ref{fig:hotspots} shows the distribution of positive hotspots, with areas of pronounced concentration, whereas others display moderate dispersion and regions devoid of fire activity are also observed. 

\subsection{Data Preprocessing and Integration}
To cope with the limited resources and computational capacity, we train our models on a subset of 500k data points, starting the selection from February 2016 due to Sentinel-3 data availability. These points are selected performing a filter on the timestamp, then a stratified sampling approach is employed to maintain the original dataset's geographical distribution and label proportions. Given the imbalanced proportion of positive and negative instances, amounting to $3\%$ and $97\%$ respectively, we undersample the negative class targeting a less skewed distribution, namely $10\%$ of positive examples with $90\%$ of negative examples, while trying to ensure the geographical distribution balancing. 
In addition to the other data sources (i.e., LULC, S3) we compute the Number of Previous Hotspots (NPH) occurring in the same place 12, 24 and 36 hours before, as well as time-related features that provide insights into the weekly periodicity of the event (\textit{Time} features). Among the available MODIS and VIIRS features, we select the Fire Radiative Power (FRP) and the following bands: t\_21, t\_31, t\_m13, t\_m15, t\_i4, t\_i5. We keep all the bands from Sentinel-3 and Land Cover.

\section{Methodology}
\label{sec:methodology}

\par\textbf{Baselines}. First, we conduct our experiments on classical machine learning algorithms, namely Logistic Regression (LR), Multi Layer Perceptron (MLP), and XGBoost (XGB). These algorithms are trained using all the computed features associated with the hotspot points. The algorithm yielding the best result acts as baseline for the subsequent experiments.

\par\textbf{RN-Van: Vanilla ResNet}. We want to utilize the complete information available from Sentinel-3 and LULC data, which covers an area of $90 km^2$ centered around the hotspot point. For this purpose, we exploit standard computer vision models to efficiently encode the information into denser, semantically meaningful, vector representation. Specifically, in this work we employ a ResNet-18 \cite{RESNET} on the S3 inputs.

\par\textbf{RN-E2E: End-To-End ResNet}. To assess the potential contribution of additional data sources, we evaluate the performance of training the ResNet-18 with a downstream Multi Layer Perceptron. The latter  takes as input the concatenation of Sentinel-3 and LULC features embedding provided by ResNet-18 and the ones from MODIS, VIIRS, Time and NPH, as shown in Figure \ref{fig:E2E}.

\section{Experiments}
\label{sec:experiments}

\subsection{Implementation details}
In our experiments, we define six Feature Sets (FS) as outlined in Table \ref{tab:featsets} in order to analyze the performance of different scenarios with varying data availability. For the experiments with ResNet, we use maps with dimension of 32x32 pixels and 33 channels (32 from Sentinel-3 and 1 from LULC).

\begin{table}[!ht]
    \centering
    \begin{tabular}{|l|c|c|c|c|c|c|}
    \hline
         & \textbf{FS1} & \textbf{FS2} & \textbf{FS3} & \textbf{FS4} & \textbf{FS5} & \textbf{FS6} \\ \hline
        \textbf{MODIS} & \checkmark & \checkmark & \checkmark & \checkmark & ~ & ~ \\ \hline
        \textbf{VIIRS} & \checkmark & \checkmark & \checkmark & \checkmark & ~ & ~ \\ \hline
        \textbf{Time} & \checkmark & \checkmark & \checkmark & \checkmark & \checkmark & ~ \\ \hline
        \textbf{Land Cover} & ~ & \checkmark & \checkmark & \checkmark & \checkmark & \checkmark \\ \hline
        \textbf{Sentinel-3} & ~ & ~ & \checkmark & \checkmark & \checkmark & \checkmark \\ \hline
        \textbf{NPH} & ~ & ~ & ~ & \checkmark & \checkmark & ~ \\ \hline
    \end{tabular}
    \caption{\label{tab:featsets}Feature Sets configuration.}
\end{table}

We divide the dataset into 50 splits following the methodology described in Section \ref{sec:dataset}. We then select 28 splits for train, 14 for validation and 8 for tests, resulting in the following ratios: 56\%, 28\%, 16\%, respectively.

Using a grid search approach, we select the optimal hyperparameters for the XGBoost model, specifically a learning rate of $\lambda = 0.1$, a tree maximum depth of 12, a feature subsample ratio of $0.8$, and a positive class weight equal to $10$.

For experiments with RN-Van and RN-E2E, we only consider the FS6 and FS4 configurations, namely the feature sets with every source included in the first case, and with only S3 and LULC in the second case. All the training processes are conducted for 20 epochs, using a batch size of 128, a learning rate of $1e^-3$, using AdamW as optimizer, and Binary Cross Entropy (BCE) Loss function. While baselines are trained on CPU only, the ResNet experiments require around 20 hours on a single NVIDIA A100 GPU.

\subsection{Results}

\begin{table}[!ht]
    \centering
    \begin{tabular}{|c|c|c|c|}
    \hline
        ~ & \textbf{LR} & \textbf{MLP} & \textbf{XGB} (baseline) \\ \hline
        \textbf{FS1} & 6.80 & 34.51 & \textbf{46.88}  \\ \hline
    \end{tabular}
     \caption{\label{tab:baseline_comparison}Comparison of baseline models in terms of F1 score}
\end{table}

\begin{table}[!ht]
\centering
\begin{adjustbox}{width=1.0\columnwidth}
    \centering
    \begin{tabular}{|c|c|c|c|cc|c|cc|}
    \hline
        \textbf{Features Set} &  FS1 &  FS2 & FS3 & \multicolumn{2}{c|}{FS4} & FS5 & \multicolumn{2}{c|}{FS6} \\ \hline
        \textbf{Model} & \textit{XGB} & \textit{XGB} & \textit{XGB} & \textit{XGB} & \textit{RN-E2E} & \textit{XGB} & \textit{XGB} & \textit{RN-Van}  \\ \hline
        \textbf{F1} &  46.88 & 56.21 & 78.99 & \textbf{80.53} & 76.47 & 80.02 & 71.40 & \textbf{76.16} \\ \hline

    \end{tabular}

\end{adjustbox}
\caption{\label{tab:comparison}Results of XGBoost (XGB) and ResNet architectures in terms of F1 score.}
\end{table}

In every experiment, we adopt the \textit{F1 score} as performance metric.
Table \ref{tab:baseline_comparison} highlights the results obtained by the baseline approaches. We observe that simple solutions such as the linear regression are not sufficient to tackle the task at hand, while the even a MLP is heavily outperformed by XGBoost. For this reason, we select the latter as baseline for the following experiments.

As evidenced by the results presented in Table \ref{tab:comparison}, expanding the feature set with additional data sources proves to be beneficial for the task.
Specifically, XGBoost manages to reach an F1 score of 80.53 in the FS4 configuration, resulting in a $+33.65$ performance improvement with respect to the baseline. Furthermore, the scores achieved with the FS5 and FS6 configurations indicate that the proposed system exhibits robust performance even in the absence of MODIS and VIIRS data sources. This capability enables the system to be deployed in near real-time scenarios, conditioned by the availability of Sentinel-3 data.
Finally, we observe that RN-Van outperforms XGBoost on FS6 with a $+4.76$ performance increase, as it likely leverages the data from the entire area more effectively, unlike XGBoost which only considers individual pixels. On the other hand, using the RN-E2E architecture incorporating all the point features with FS4, leads to a negligible improvement compared to RN-Van (i.e., $+0.31$). Similarly, this can probably be attributed to the underlying architecture that cannot exploit the external features as well as the visual ones. XGBoost still demonstrates superior results in this configuration.

\section{Conclusions and future works}
\label{sec:conclusions}

We proposed a novel multimodal supervised machine learning approach to enable a better automated near real-time wildfire detection system. The proposed approach process data from multiple satellite sources, showing that by adding more characterizing data sources, the wildfire detection performance increases. Moreover, promising results can also be obtained using data from Sentinel-3 and LULC only. 

Future research endeavours could encompass the extension of the applicability of this work to address diverse fire types, including small ones, or focus on a more precise real-time fire detection system by leveraging time series analysis on geostationary satellites such as the upcoming Meteosat generation three.

\bibliographystyle{IEEEbib}
{\footnotesize
\bibliography{strings.bib,refs.bib}

\begin{thebibliography}{1}

\bibitem{ipcc2022report}
The Intergovernmental~Panel on~Climate Change~(IPCC),
\newblock ``{Climate Change 2022: Impacts, Adaptation and Vulnerability},''
  2022,
\newblock [Online; accessed 17-May-2023].

\bibitem{Giglio2003AnEC}
Louis Giglio, Jacques Descloitres, Christopher~O. Justice, and Yoram~J.
  Kaufman,
\newblock ``An enhanced contextual fire detection algorithm for modis,''
\newblock {\em Remote Sensing of Environment}, vol. 87, pp. 273--282, 2003.

\bibitem{GiglioVIIRS}
Wilfrid Schroeder, Patricia Oliva, Louis Giglio, and Ivan~A. Csiszar,
\newblock ``The new viirs 375m active fire detection data product: Algorithm
  description and initial assessment,''
\newblock {\em Remote Sensing of Environment}, vol. 143, pp. 85--96, 2014.

\bibitem{schroeder2016}
W~Schroeder, I~Csiszarb, L~Giglioc, E~Ellicottc, and C~Justicec,
\newblock ``Satellite active fire product validation using high spatial
  resolution reference data,''
\newblock .

\bibitem{engel2020}
Chermelle~B. Engel, Simon~D. Jones, and Karin Reinke,
\newblock ``A seasonal-window ensemble-based thresholding technique used to
  detect active fires in geostationary remotely sensed data,''
\newblock {\em IEEE Transactions on Geoscience and Remote Sensing}, vol. 59,
  no. 6, pp. 4947--4956, 2021.

\bibitem{fu2020}
Yuyun Fu, Rui Li, Xuewen Wang, Yves Bergeron, Osvaldo Valeria, Raphaël~D.
  Chavardès, Yipu Wang, and Jiheng Hu,
\newblock ``Fire detection and fire radiative power in forests and low-biomass
  lands in northeast asia: Modis versus viirs fire products,''
\newblock {\em Remote Sensing}, vol. 12, no. 18, 2020.

\bibitem{coskuner2022}
Kadir~Alperen Coskuner,
\newblock ``Assessing the performance of modis and viirs active fire products
  in the monitoring of wildfires: a case study in turkey,''
\newblock {\em iForest - Biogeosciences and Forestry}, , no. 2, pp. 85--94,
  2022.

\bibitem{9553499}
Krishna Karra, Caitlin Kontgis, Zoe Statman-Weil, Joseph~C. Mazzariello, Mark
  Mathis, and Steven~P. Brumby,
\newblock ``Global land use / land cover with sentinel 2 and deep learning,''
\newblock in {\em 2021 IEEE International Geoscience and Remote Sensing
  Symposium IGARSS}, 2021, pp. 4704--4707.

\bibitem{RESNET}
Kaiming He, Xiangyu Zhang, Shaoqing Ren, and Jian Sun,
\newblock ``Deep residual learning for image recognition,''
\newblock {\em CoRR}, vol. abs/1512.03385, 2015.

\end{thebibliography}
}
\end{document}